\documentclass[amssymb,aps,showpacs,showkeys]{revtex4}

\usepackage{graphicx}
\usepackage{bm}

\begin{document}

\title{Synthesis of Taylor Phase Screens with Karhunen-Lo\`eve Basis Functions}

\author{Richard J. Mathar}
\homepage{http://www.strw.leidenuniv.nl/~mathar}
\email{mathar@strw.leidenuniv.nl}
\affiliation{Leiden Observatory, Leiden University, P.O. Box 9513, 2300 RA Leiden, The Netherlands}


\date{\today}

\begin{abstract}
Phase screens above a telescope pupil represent the variation of the 
phase of the electromagnetic field induced by atmospheric turbulence.
Instances drawn from such statistics are represented by a vector of
random phase amplitudes which are coefficients of a linear superposition
of two-dimensional basis functions across the pupil. This work shortly reviews
Fried's analysis of this modal decomposition for the case of Kolmogorov
statistics of the phase covariance as a function of separation in the
pupil plane.

We focus on the numerical synthesis of phase screens. The statistically
independent modes are transformed into the eigen-modes of a gradient matrix
as time-dependence is introduced such that on short time scales the
instances of the phase screens are rigidly shifted into a direction
imposed by some wind velocity---known as the Taylor frozen screen
approximation. This simple technique factorizes spatial and temporal
variables and aims at binding the time dependence of the phase
screens to the few expansion coefficients of the basis functions that
obey a stochastic time-dependent differential equation.
\end{abstract}

\maketitle
\section{Overview}
\subsection{Phase Screen Snapshots}

A phase screen summarizes the phase delays in sub-apertures of
a telescope of some diameter $D$ on lateral spatial scales that
depend on the inhomogeneity of the refractive index along the path
through the few kilometers of atmosphere in ground-based observations
\cite{HufnagelJOSA54}.

The intent to remove the time-dependent sparkling of star light
in (almost all) applications of astronomy leads to adaptive optics
and is increasingly important as telescope diameters increase and/or
anisoplanatism puts limits on coherency in wide-angle observations
\cite{WhiteleyJOSAA15,ConanJOSA17,RoddierProgOpt19}.
The equivalent
variables in long baseline interferometry spawn interest in
the definition of outer scales.

The impact on imaging would be determined by the degrees of
freedom and stiffness of response in any subsequent 
adaptive optics correction, and by
additional weighting with spatial filter functions as
found in some fiber conductive followup systems
\cite{KeenMNRAS326,ShaklanAO27,MennessonJOSA19,WallnerJOSA19}.

This work is a contribution to the numerical simulation
of phases $\varphi({\bf r},t)$ in some pupil plane taking
a two-dimensional spatial vector ${\bf r}$ and a time $t$ as arguments.
We follow Fried's analysis of the build-up of the phase
at some snapshot in time if the covariance (or structure function)
follows a Kolmogorov power-law as a function of the distance between
two points in $|{\bf r}|\le D/2$ 
\cite{FriedJOSA68,BarenblattSIR40}. Assuming that the number and size of the
speckles is founded on Gaussian statistics after a sufficiently
long path through the atmosphere, the phase can be synthesized
by linear superposition of two-dimensional basis functions multiplied by a
vector of random variables with individual Gaussian statistics.
The statistically independent basis functions are constructed
as eigenfunctions of the generic Karhunen-Lo\`eve (KL) integral equations
which incorporate isotropy and strength of the covariance in the
integral kernel and the sampling limits of the statistics (here: the
circular pupil) in the region of integration.

\subsection{Taylor Movies}

With some assumption of ergodicity, independent snapshots of
phase screens are generated by throwing dice for each new set of expansion
coefficients with a random number generator. On short time scales,
however, the speckle boiling time (coherence time) would be
overtaken by the assumption that the phase screen would move
---in the fixed pupil coordinates---
rigidly and predictably as drawn by some wind velocity $v$ \cite{PoyneerJOSAA,SchwartzJOSAA11}.
The expansion coefficients follow
a stochastic Ornstein-Uhlenbeck-type
differential equation which ensures deterministic (first-order
smooth) alignment of the pictures of the Taylor movie, but allows
transformation of each expansion coefficient within the amplitudes
set by the eigenvalue of its KL mode.

The alternatives to this ansatz are the generation of static
phase screens much larger than the pupil diameter, from which
a moving disk is cut off to define the region
of interest \cite{AssematOE14,JakobssonAO35,SedmakAO43,BrummelaarOptComm132,VorontsovWRM18,FriedJOSAA25,DiosOE16,MatharWRCM19}.
The principle
of embedding the pupil into a larger virtual phase screen will also
be the guideline to define on which timescale how much randomness
will be mixed into the expansion coefficients. The main difference
is the efficiency of maintaining either a small set of time-dependent
expansion coefficients with the option to generate movies
of arbitrary duration (here), or maintaining a spatial stripe of the sky in computer memory
which has the width of the pupil but a much larger length
determined by the product of movie duration and wind speed.

A summary of the well-known algebra of KL eigenmodes is
provided in section \ref{sec.KL}\@.
The differential equation and dynamic matrix of the motion
implied by the Taylor frozen screen assumption is written down
and discussed in section \ref{sec.tayl}, and notes of expanding
this into a stochastic equation follow in section \ref{sec.stoch}\@.
Some of the mathematical aspects are
outsourced to the appendix.

\section{KL Eigenfunctions of Kolmogorov Covariance} \label{sec.KL} 
\subsection{Separation in Radial and Azimuthal Functions}

A summary of the established theory and mathematics of phase patterns
\cite{DaiJOSAA12,WangJOSA68,FriedJOSA68,RoddierOE29}
is given in this section, accompanied by errata in Appendix \ref{app.err} 
for
the key papers.
The phase of the electromagnetic field in the pupil plane is decomposed
into basis functions $F({\bf r})$ and fluctuating expansion coefficients $a$
\begin{equation}
\varphi({\bf r}) = \sum_j a_j F_j({\bf r}), \quad |{\bf r}|<D/2.
\label{eq.ai}
\end{equation}
If the $a$ are supposed to build a vector of independent scalar variables which vary
from one phase screen to another, the covariance statistics between two
pickup positions of the phase translates into a requirement of the
basis functions, which are forced to become KL eigenfunctions
of a linear integral
operator. If the two-dimensional region of sampling the phase statistics
matches the symmetry in the covariance matrix, a separation ansatz within
the $F_i$ is fruitful. In particular, if the domain of the pupil is circular
and the covariance isotropic (only depending on the separation $|{\bf r}-{\bf r'}|$),
the eigenfunctions can be separated in azimuthal and radial terms,
\begin{equation}
F_j({\bf r})=K_p^{(q)}(x) M_q(\theta),
\quad 0\le x,
\quad 0\le \theta\le 2\pi.
\end{equation}
If the phase structure function $\cal D$ obeys a power law
\begin{equation}
{\cal D}=2c_\varphi (|{\bf r}-{\bf r}'|/r_0)^{1+\gamma}, \quad \gamma=2/3,
\label{eq.powlaw}
\end{equation}
one can essentially reduce the KL integral equation to a universal
form by decoupling the scaling parameter $D/r_0$ from the $F_j$ and moving
it into the coefficients $a_j$. The scale factor is \cite{FriedJOSA56_1372,RoddierProgOpt19}
\begin{equation}
2c_\varphi
=2\left[4\Gamma\left(\frac{3+\gamma}{1+\gamma}\right)\right]^{(1+\gamma)/2}
=2[(24/5)\Gamma(6/5)]^{5/6}\approx 6.883877.
\end{equation}
The Fried parameter $r_0$ is essentially proportional to the $6/5$-th power
of the observing wavelength \cite{FriedJOSA68} if the wavelength dependence
of the structure constant of the refractive index remains negligible.
I shall work with basis functions which are normalized over the scaled
distance $x= 2 \openone_x |{\bf r}|/D$ from the pupil center,
\begin{equation}
\int_D F_j^2({\bf r}) d^2r =
\int_0^{\openone_x} x {K_p^{(q)}}^2(x) dx =
\int_0^{2\pi} M_q^2(\theta)d\theta=1.
\label{eq.normK}
\end{equation}
The constant $\openone_x$ is set to $1/2$ if radial distances are measured in units of the
pupil diameter $D$, and set to $1$ if radial distances are measured in units of 
the pupil radius $D/2$. The only purpose of this notation is to keep track of
both choices that have been in use
in the literature.
The azimuthal basis functions
\begin{equation}
M_q(\theta)=\sqrt{\frac{\epsilon_q}{2\pi}}\times \left\{
\begin{array}{c}
\cos (q\theta), \\
\sin (q\theta), \\
\end{array}
\right.
\label{eq.Mtheta}
\end{equation}
will be labeled with positive $q$ for the cosine type and negative $q$
for the sine type, akin to the nomenclature for the two
symmetries of Zernike polynomials \cite{NollJOSA66}.
The Neumann factor $\epsilon$ is defined as
in the
literature on Bessel Functions,
\begin{equation}
\epsilon_q\equiv \left\{
\begin{array}{ll}
1, & q=0 ,\\
2, & |q|\ge 1 .\\
\end{array}
\right.
\end{equation}
The radial eigenmodes $K_p^{(q)}$ for each specific azimuthal ``quantum''
number $q$ are calculated as eigenvectors of the KL equation
\cite[(25)]{FriedJOSA68}
\begin{equation}
\frac{1}{2\openone_x}\int_0^{\openone_x} R_q(x,x')K_p^{(q)}(x')dx'=\lambda_{p,q}^2 K_p^{(q)}(x)
\label{eq.kh}
\end{equation}
with eigenvalues $\lambda_{p,q}^2$. The integral kernel is given by
\begin{eqnarray}
2\openone_x R_0(x,x')&=& -c_\varphi x'\int_0^{2\pi}
\frac{1}{(2\openone_x)^{1+\gamma}}
(x^2+x'^2-2xx'\cos\theta')^{(1+\gamma)/2} d\theta'
+2\pi x'[{\cal G}_1(x)+{\cal G}_1(x')-{\cal G}_2]
,
\label{eq.R0}
\\
2\openone_x R_q(x,x')&=& -c_\varphi x'\int_0^{2\pi}
\frac{1}{(2\openone_x)^{1+\gamma}}
(x^2+x'^2-2xx'\cos\theta')^{(1+\gamma)/2} \cos(q\theta') d\theta',\quad q\neq 0,
\end{eqnarray}
where
\begin{eqnarray}
{\cal G}_1(x)&=& c_\varphi \frac{4}{\pi(2\openone_x)^{3+\gamma}}\int_0^{\openone_x}dx'' x''
\int_0^{2\pi}(x^2+x''^2-2xx''\cos\theta'')^{(1+\gamma)/2} d\theta'', \\
{\cal G}_2&=& \frac{8}{(2\openone_x)^2}\int_0^{\openone_x}dx'' x'' {\cal G}_1(x'').
\label{eq.G2}
\end{eqnarray}
Since we do not assume that the expectation value of the tip-tilt component
of the phase over the pupil vanishes, the terms proportional to
Fried's variables ${\cal G}_3$ and ${\cal G}_4$ do not appear in the
covariance and vanish in our analysis. Consequently, here and
in \cite{WangJOSA68,DaiJOSAA12}, tip-tilt modes are in the list
of KL eigenfunctions,
but not in
Fried's list.

\subsection{Implementation}
In numerical practise, the KL equation is solved for the symmetrized variables,
here marked with a tilde,
\begin{eqnarray}
\tilde K_p^{(q)}(x)\equiv \sqrt{\frac{x}{2\openone_x}} K_p^{(q)}(x), \\
R_q(x,x')\equiv \frac{x'}{2\openone_x} \tilde R_q(x,x'),
\quad
\tilde R_q(x,x')=\tilde R_q(x',x),
\end{eqnarray}
which turn (\ref{eq.kh}) into
\begin{equation}
\frac{1}{(2\openone_x)^2}\int_0^{\openone_x} \sqrt{xx'}\tilde R_q(x,x')\tilde K_p^{(q)}(x')dx'
=\lambda_{p,q}^2 \tilde K_p^{(q)}(x).
\label{eq.khsymm}
\end{equation}
The benefits of working with an integral kernel that is symmetric
under the exchange $x\leftrightarrow x'$ are
\begin{itemize}
\item
numerical stability and performance by use of linear algebra
eigenvalue solvers for this type of symmetry.
\item
immediate evidence that the eigenvalues are real-valued with orthogonal
eigenvectors,
\begin{eqnarray}
\int_0^{\openone_x} \tilde K_p^{(q)}(x) \tilde K_{p'}^{(q)}(x) dx &=& \delta_{pp'}, \\
\int_0^{\openone_x} x K_p^{(q)}(x) K_{p'}^{(q)}(x) dx &=& \delta_{pp'}.
\label{eq.ortho}
\end{eqnarray}
\end{itemize}

Further comments on the numerical treatment are given in
Appendix \ref{app.hyper}. Some variants of the actual
representation of the radial functions $K_p^{(q)}$ exist.
A simple and stable format is the finite-element (FEM) representation, in which
$\tilde K_p^{(q)}$ is a vector of values on (possibly equidistant)
$x_i$, $i=1,2,\ldots,N$. In this case the matrix representation of the kernel
of the KL equation (\ref{eq.khsymm}) is the $N\times N$ table of
the gridded $\sqrt{xx'}\tilde R_q(x,x')$ multiplied by any weights
associated with the numerical integration.
A power basis
\begin{equation}
\tilde K_p^{(q)}(x)=x^{1/2+|q|} \sum_{j=0}^\infty k_{jpq}x^j
\label{eq.Kofxpow}
\end{equation}
has the (small) disadvantage that---after insertion into (\ref{eq.khsymm}) and
out-projection of the $k_{ipq}$---a non-diagonal overlap matrix is left
on the right hand side which leads to a slightly more complicated
generalized eigenvalue problem. This is easily avoided by moving on
to a Zernike basis of orthogonal Jacobi Polynomials.

Figures of the basis functions $F(\bm{r})$ with the largest eigenvalues,
which are dominant and represent the speckles of largest size have
been shown before \cite[Fig.\ 5b]{WangJOSA68}.
The order of the rotation
axis is determined by $|q|$; the eigenmodes show up in pairs that
can be mutually converted by rotation around the pupil center by
angles of $\pi/(2q)$, as established by $M_q(\theta)$.

\subsection{Wave number (Fourier) Representation} \label{sec.Kofsig}
The two-dimensional Fourier transform of the reduced KL eigenfunctions $K_p^{(q)}M_q$ is
\begin{eqnarray}
F_j(\bm{\sigma}) &=&  \int_{x\le \openone_x} d^2x K_p^{(q)}(x)M_q(\theta) \exp\left(2\pi i \bm{\sigma}\cdot {\bf x}\right)
=\int_0^{\openone_x}xdx\int_0^{2\pi}d\theta K_p^{(q)}(x)M_q(\theta)\exp[2\pi i\sigma x \cos(\theta_\sigma-\theta)]
\nonumber \\
&=& K_p^{(q)}(\sigma)M_q(\theta_\sigma)
,
\label{eq.Fofsigma}
\end{eqnarray}
where $\sigma\equiv |\bm{\sigma}|$ and $\theta_\sigma$ define
the spherical coordinates of the wave number, and
\begin{equation}
K_p^{(q)}(\sigma)\equiv
2\pi i^q\int_0^{\openone_x}x K_p^{(q)}(x)J_q(2\pi\sigma x)dx;
\quad
K_p^{(q)}(x)\equiv
2\pi (-i)^q\int_0^{\infty}\sigma K_p^{(q)}(\sigma)J_q(2\pi\sigma x)d\sigma
\label{eq.Kofsigma}
\end{equation}
is a Fourier pair.
Not to insert
factors of $\openone_x$ in the definition of the Fourier transform here
is a judicious choice to ensure
that the normalization (\ref{eq.ortho}) is the same in the $x$ and in the $\sigma$
domain:
\begin{equation}
\int_0^\infty \sigma K_p^{(q)}(\sigma)K_{p'}^{(q)*}(\sigma)d\sigma = \delta_{pp'}.
\label{eq.KorthoF}
\end{equation}

If the $K_p^{(q)}(x)$ are expanded in a series of Zernike polynomials
(App.\ \ref{zern.app}), the $K_p^{(q)}(\sigma)$ are the equivalent series
of Bessel Functions \cite[(8)]{NollJOSA66}\cite{Born}.

\section{Taylor Model} \label{sec.tayl} 
\subsection{Equation-of-Motion and Gradient Matrix}

The theme of this paper is how any two of the sets of coefficients $a_j$
are morphed if time is a added as a new parameter to the description.

To ensure steady transformation in short time intervals, we will employ
the Taylor model of lateral displacement into the direction of a velocity
vector ${\bf v}$, which is represented by \cite{ConanJOSA12}
\begin{equation}
\varphi({\bf r},t)
=
\varphi({\bf r}-{\bf v} t,0)
\label{eq.tayl}
\end{equation}
for time $t$, distance $r$ to the pupil center and azimuth $\theta$ along
the pupil periphery \cite{RoddierJOSAA10}.
We make the scaling of the coefficients $a_j$ by the eigenvalues
$\lambda$ explicit by writing the expansion as
\begin{equation}
\varphi(r,\theta,t)
=
\sum_{l,m} \beta_l^{(m)}(t) \lambda_{l,m} K_l^{(m)}(x)
M_m(\theta),\quad x=2\openone_x r/D.
\label{eq.phishif}
\end{equation}
The fundamental scaling parameters
$(D/r_0)^{1+\gamma}$ and $\lambda$
can all be partially absorbed in these factors or basis functions. The $a_j$
in (\ref{eq.ai}) obey a statistics with variances $(D/r_0)^{1+\gamma}\lambda_i^2$,
so the components of the random vector $\beta_i^{(m)}$ introduced here have
all the same variance, $(D/r_0)^{1+\gamma}$.
The Taylor model is basically a means to substitute the time derivative
in the equation-of-motion (EOM) of $\varphi$ by a gradient,
$\partial_t\rightarrow -{\bf v}\cdot \nabla_{\bf r}$,
\begin{equation}
\partial_t\varphi({\bf r},t)
=
\sum_{l,m} \frac{\partial\beta_l^{(m)}(t)}{\partial t} \lambda_{l,m} K_l^{(m)}(r)
M_m(\theta)
=
-{\bf v}\cdot \nabla_{\bf r}\varphi({\bf r},0)
.
\label{eq.ddtphi}
\end{equation}
This is the infinitesimal version of de-centering the basis
functions \cite{ComastriJOpt9,LundstromJOSAA24,HerrmannJOSA71}.
The corresponding requirement in Fourier space is
\begin{equation}
\varphi(\nu,\bm{\sigma})=\varphi(\bm{\sigma})\delta(\nu-\bm{\sigma}\cdot {\bf v})
.
\label{eq.four}
\end{equation}
In polar coordinates
in the pupil plane,
\begin{equation}
\theta=\arctan\frac{Y}{X};\quad r=\sqrt{X^2+Y^2},
\end{equation}
the two components of the gradient operator are \cite{ChurnsideOL10,BrummelaarOptComm115}
\begin{eqnarray}
\partial_X
&=&
\frac{\partial \theta}{\partial X}\partial_\theta
+\frac{\partial r}{\partial X}\partial_r
=-\frac{\sin \theta}{r}\partial_\theta
+\cos \theta\partial_r,
\\
\partial_Y &=& 
\frac{\partial \theta}{\partial Y}\partial_\theta
+\frac{\partial r}{\partial Y}\partial_r
=\frac{\cos \theta}{r}\partial_\theta
+\sin \theta\partial_r.
\label{eq.ddypolar}
\end{eqnarray}
To tighten the notation, we assume that the velocity has no component
in the $Y$-direction of coordinates, 
so (\ref{eq.ddypolar}) is not needed and (\ref{eq.ddtphi}) becomes
\begin{eqnarray}
&&
\sum_{l,m} \frac{d\beta_l^{(m)}(t)}{dt}\lambda_{l,m} K_l^{(m)}(r)
M_m(\theta)
=
-v\partial_X\varphi({\bf r},0)
=
v\left(\frac{\sin\theta}{r}\partial_\theta-\cos\theta\partial_r\right)
\sum_{j,n} \beta_j^{(n)}(t) \lambda_{j,n} K_j^{(n)}(r)
M_n(\theta) \\
&=&
v
\sum_{j,n} \beta_j^{(n)}(t) \frac{\lambda_{j,n}K_j^{(n)}(r)}{r}\times
\frac{\sqrt\epsilon_n}{\sqrt{2\pi}}
\left\{ \begin{array}{c} -n\sin\theta\sin n\theta \\ n\sin\theta\cos n\theta \end{array}\right.
-v
\sum_{j,n} \beta_j^{(n)}(t)\frac{\lambda_{j,n}\partial K_j^{(n)}(r)}{\partial r}
\times
\frac{\sqrt\epsilon_n}{\sqrt{2\pi}}
\left\{ \begin{array}{c} \cos\theta\cos n\theta \\ \cos\theta\sin n\theta \end{array}\right.
\\
&=&
\frac{v}{2}
\sum_{j,n>0}n \beta_j^{(n)}(t) \frac{\lambda_{j,n}K_j^{(n)}(r)}{r}\times
\frac{\sqrt\epsilon_n}{\sqrt{2\pi}}
\left\{ \begin{array}{c}  \cos (n+1)\theta -\cos (n-1)\theta \\
                          \sin (n+1) \theta -\sin (n-1) \theta
\end{array}
\right.
\nonumber \\
&&
-\frac{v}{2}
\sum_{j,n} \beta_j^{(n)}(t)\frac{\lambda_{j,n}\partial K_j^{(n)}(r)}{\partial r}
\times
\frac{\sqrt\epsilon_n}{\sqrt{2\pi}}
\left\{ \begin{array}{c}
\cos (n+1)\theta+\cos (n-1)\theta \\ \sin (n+1)\theta+\sin (n-1)\theta \end{array}\right.
.
\label{eq.gradLadd}
\end{eqnarray}
This write-up is a composite of an upper line for the even and a lower
line for the odd $m$.
The upper line refers to
the $M_m$ cosine modes, including the radially symmetric $m=0$ modes, and the
lower line refers to the $M_m$ sine modes.
The coupling is between azimuthal parameters that differ by one,
$m\leftrightarrow n\pm 1$, similar to the selection rules of the
electric dipole operator between hydrogenic states of the Schr\"odinger atom.

\subsection{Hybridization of KL Eigenmodes}

To isolate one coefficient, we multiply (\ref{eq.gradLadd})
by a general $\lambda_{k,s}K_k^{(s)}(x)M_k(\theta)$, substitute
$r\rightarrow xD/(2\openone_x)$, and integrate over the pupil, exploiting
the orthogonality relations (\ref{eq.ortho}):
\begin{eqnarray}
\frac{D}{v\openone_x}\frac{d}{dt} \beta_k^{(s)}
&=&
\frac{1}{\lambda_{k,s}^2}\big[
\sum_{j,n>0} n\beta_j^{(n)}
\int_0^{\openone_x}dx \lambda_{j,n}K_j^{(n)} \lambda_{k,s}K_k^{(s)}
\frac{\sqrt{\epsilon_n\epsilon_s}}{\epsilon_s}
\delta_{n+1,s}
\nonumber \\
&&
-
\sum_{j,n>0} n\beta_j^{(n)}
\int_0^{\openone_x}dx \lambda_{j,n} K_j^{(n)} \lambda_{k,s}K_k^{(s)}\frac{\sqrt{\epsilon_n\epsilon_s}}{\epsilon_s}
\delta_{|n-1|,s}
\nonumber \\
&&
-
\sum_{j,n\ge 0} \beta_j^{(n)}
\int_0^{\openone_x}xdx \lambda_{j,n} \partial_x K_j^{(n)} \lambda_{k,s}K_k^{(s)}\frac{\sqrt{\epsilon_n\epsilon_s}}{\epsilon_s}
\delta_{n+1,s}
\nonumber \\
&&
-
\sum_{j,n\ge 0} \beta_j^{(n)}
\int_0^{\openone_x}xdx \lambda_{j,n} \partial_x K_j^{(n)} \lambda_{k,s}K_k^{(s)}\frac{\sqrt{\epsilon_n\epsilon_s}}{\epsilon_s}
\delta_{|n-1|,s}
\big],\quad s\neq 0 .
\end{eqnarray}
This is a system of linear homogeneous differential equations with a skew-symmetric,
real-valued, sparse matrix $\Omega$,
\begin{equation}
\frac{D}{2 v}\frac{d}{dt} \beta_k^{(s)}
=\sum_{nj}\Omega_{ks,jn} \beta_j^{(n)} ,
\quad
\Omega^T=\Omega^\dagger=-\Omega,
\label{eq.EOM}
\end{equation}
where the symbols $T$ and $\dagger$ mean transposition and Hermite
conjugation, respectively. Computation of the matrix elements in
Fourier space is proposed in Appendix \ref{sec.OmegaFT}\@.
Existence of a skew-symmetric representation is expected
from the fact that the gradient operator changes sign with the parity
of a polar vector. In
the Cartesian $X$-$Y$ coordinate system, it is coupled to the sign change
of the derivative integrals after partial integration (assuming the
``surface'' integrals of Green's law vanish), but it is less
obvious for the set of integrals reduced to the radial coordinates $r$
or $x$ and assembled in $\Omega$. The analytic proof works
with the derivative of Mercer's theorem of the
covariance function in the KL kernel,
\begin{equation}
\frac{1}{2\openone_x} R_q(x,x')=\sum_p \lambda_{p,q}^2K_p^{(q)}(x) K_p^{(q)}(x')
\end{equation}
and is omitted here. It is mainly to retain this symmetry feature of the
gradient matrix
that I chose to split off the values of $\lambda$ in (\ref{eq.ddtphi}) instead of
working with the $a_j$ expansion coefficients.

Skew-symmetry implies that the  eigenvalues of the 
transformation to principal axes with some orthogonal matrix $\xi$,
\begin{equation}
\Omega = \xi i\omega_l \xi^\dagger
;\quad
\xi\xi^\dagger=1,
\end{equation}
are pairs of complex-conjugate, purely imaginary numbers $i\omega_l$
\cite{ThompsonPAMS104,PardekooperNumMath17,BunchMathComp38,BennerETNA11,KressnerBIT}.
This transformation of the basis of the $\beta$ vector induces 
a transformation of the KL basis,
\begin{equation}
\varphi =
\sum_{j,m} \left(\beta_j^{(m)}\right)^T \lambda_{j,m}K_j^{(m)}M_m
=
\left(\xi^*\beta_j^{(m)}\right)^T \xi \lambda_{j,m}K_j^{(m)}M_m
.
\end{equation}
Whereas the KL eigenfunctions
have a rotational symmetry in the pupil plane inherited from $M(\theta)$,
each gradient eigenfunction shows some undulation along the
$v$ direction (here: the direction of $X$) at a
spatial period
of $\pi D/(\omega_l)$.

From the mathematical point of view, the
standard aim of this basis transformation
is to decouple the EOM's
of the expansion coefficients
(\ref{eq.EOM}),
\begin{equation}
\frac{D}{2}\frac{d(\xi^*\beta_k^{(s)})_l}{v\, dt}
= i\omega_l(\xi^*\beta_k^{(s)})_l,
\end{equation}
which solves the time-dependent differential equation in terms of
oscillatory fluctuations from initial values,
\begin{equation}
\hat \beta_l(t)
=
\exp\left(\frac{2 v}{D}i\omega_l t\right) \hat \beta_{l\vert t=0}
.
\label{eq.betahatt}
\end{equation}
where
$\hat\beta_l$ is the $l$-th component of the matrix-vector product $\xi^*\beta_k^{(s)}$.
These solutions at
discrete angular frequencies $\omega=2 v\omega_l/D$
explain in conjunction with (\ref{eq.four}) why the diagonalized eigen-modes have well-defined
spatial frequencies along the ${\bf v}$-direction.

From a less formal point of view, the basis transformation is the definition
of waves traveling in $v$ direction under the conditions of
\begin{itemize}
\item
compatibility with the structure function,
\item
enabling smooth dragging of the phase screen
(compatibility with the Taylor hypothesis)
by pairing of gradient eigenfunctions that are shifted relative
to each other along $v$ by a quarter of the spatial period.
\end{itemize}
This is to be compared with the
alternative of
starting from the structure function as a power density function
in wave number space \cite{PoyneerJOSAA19,LaneWRM2}.
A key  ingredient of the KL equation is its
dependence on the finite support
(here: the circular shape of the pupil)
which samples the covariance \cite{PrasadJOSAA16}.
The Gibb's oscillations which
represent this cutoff are a major element of the Fourier decomposition
of the KL functions.
Working in real space with the gradient matrix might be interpreted
as a deconvolution of these, followed by a superposition such that
components on ridges of a given projection
in the ${\bf v}$-direction,
i.e.\ $\sigma M_1(\theta_\sigma)=$const in our notation, are accumulated.

The phase screen is generated by using these time-dependent coefficients
as multipliers for the gradient eigenfunctions. The latter are static and
have to be generated only once for a particular set of KL functions,
\begin{equation}
\varphi({\bf r},t) 
=
\left[\exp\left(i\frac{2v}{D}\omega_l t\right) (\xi^*\beta_k^{(s)})_{\vert t=0}\right]^T
\xi \lambda_{k,s}K_k^{(s)}(r)M_s(\theta)
=
\sum_l \hat \beta_l(t)\hat F_l({\bf r})
,
\end{equation}
where
$F_l$ is the $l$-th component of the matrix-vector product $\xi\lambda_jF_j$.
The transformed initial coefficients $(\xi^* \beta)_{\vert t=0}$ are a linear superposition
of Gaussian random variables and therefore
Gaussian random
variables themselves.
Since $\xi$ is
orthogonal and since splitting off $\lambda$ in (\ref{eq.phishif}) made
the $\beta$
identically distributed (iid), the independence is
sustained by the transformation \cite{PestmanEM53}.

An accurate solution along these lines for some initial
values does not actually produce a simple shift as a function of time;
one reasons is that any finite basis set remains incomplete and
the coupling to the fine-grained modes is missing---which will be discussed
in Section \ref{sec.stoch}\@.

\section{Resumption of Stochastic Time Dependence} \label{sec.stoch} 

\begin{figure}
\includegraphics[scale=0.49]{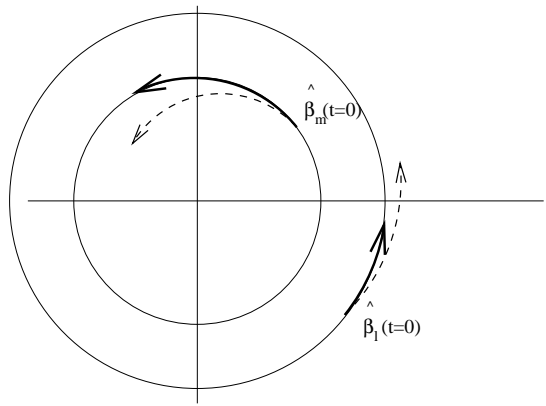}
\caption{The solution (\ref{eq.betahatt}) of the differential equation
places
the components (here: $m$ and $l$) of the vector $\hat \beta(t)$ of
expansion coefficients on circular orbits with different velocities (bold arrows).
To ensure that each component
samples its full Gaussian distribution over large time intervals, a stochastic
term needs to be added such that the simulated motion becomes chaotic
as $t$ increases (dashed circles with variable radii).
\label{fig.beta}
}
\end{figure}
The use of the Taylor hypothesis in Section \ref{sec.tayl}
puts a tangent condition on the components $\hat \beta_l(t)$
at each point in time, which lets them
wander on circles in the complex plane once some 
randomly selected $\hat \beta_{l\vert t=0}$ have been chosen.
For a finite set of basis functions, say $1\le j\le N$ in (\ref{eq.ai}),
we have essentially constructed some smooth embedding of the initial, randomized
phase screen into an infinitely large phase screen, which turns out
to be periodic in $N$ variables.
This determinism looks like an artifact if one aims at creation of realistic time
series;
actually, each component ought become independent of
(uncorrelated to) its initial value on large time scales,
which calls for some modification of the motion of the components
as
indicated in Fig.\ \ref{fig.beta}\@. This would possibly be achieved
by conversion of the deterministic differential equation (\ref{eq.ddtphi})
into some Ornstein-Uhlenbeck process by adding a time-dependent stochastic
term \cite{BeghiJOSAA25}.

We conclude
with remarks on how the Taylor screen ansatz
can be employed to dither the time-dependent expansion coefficients
without searching for such an explicit (in the nomenclature of differential equations:
inhomogeneous) term of the EOM\@.
The differential form (\ref{eq.ddtphi}) is not strictly equivalent
to the translational symmetry (\ref{eq.tayl}):
The value of $\varphi({\bf r}-{\bf v}t,0)$
on the right hand side of (\ref{eq.tayl})
is generally---after times of the order $D/v$---peeking outside the pupil of size $D$;
it is not known precisely at $t=0$ because the $N$ basis functions and
associated expansion coefficients have
only been endowed with the representation of the phase screen inside $D$.
With the Taylor hypothesis we can virtually reach
out to ${\bf r}-{\bf v}t$ if we move on to a size
\begin{equation}
\bar D>|{\bf r}-{\bf v}t|,
\label{eq.Dbaroft}
\end{equation}
embedding it such that in the inner part consistently
\begin{equation}
\varphi_D({\bf r},t)
=
\varphi_D({\bf r}-{\bf v} t,0)
=
\varphi_{\bar D}({\bf r}-{\bf v} t,0),
\end{equation}
where $\varphi_{\bar D}$ is a linear adaptive function of
$\varphi_D$.
A set
of higher order adaptive functions within a super-aperture
$\bar D>D$ would contain the full information to implement (\ref{eq.tayl}) at finite
times, and would be constructed with a basis set size $\bar N>N$ to
account for the additional degrees of freedom to represent the
speckles in the
annulus outside $D$.

For a quantitative model, one assumes that an excess $\bar N-N$ basis functions
are discarded while synthesizing the phase screen. Working with the smaller basis
set at the argument $x=r/D$, the rightmost $\bar N-N$ columns of the full
matrix transformation from the arguments $r/\bar D$ of the host
phase screen
\begin{eqnarray}
&&
\left(
\begin{array}{ccc}
K_0^{(q)}(2\openone_x r_1/D) & K_1^{(q)}(2\openone_x r_1/D) & \ldots \\
K_0^{(q)}(2\openone_x r_2/D) & K_1^{(q)}(2\openone_x r_2/D) & \ldots \\
\ldots & \ldots & \cdots
\end{array}
\right)
\nonumber
\\
&& \quad =
\left(
\begin{array}{ccc|c}
K_0^{(q)}(2\openone_x r_1/\bar D) & K_1^{(q)}(2\openone_x r_1/\bar D) & \ldots & K_k^{(q)}(2\openone_x r_1/\bar D) \ldots \\
K_0^{(q)}(2\openone_x r_2/\bar D) & K_1^{(q)}(2\openone_x r_2/\bar D) & \ldots & K_k^{(q)}(2\openone_x r_2/\bar D) \ldots \\
\ldots & \ldots & \cdots
\end{array}
\right)
\cdot
\left(
\begin{array}{ccc}
\Gamma_{00}^{(q)} & \Gamma_{10}^{(q)} & \ldots \\
\Gamma_{01}^{(q)} & \Gamma_{11}^{(q)} & \ldots \\
\ldots & \ldots & \cdots \\
\hline \\
\Gamma_{0k}^{(q)} & \Gamma_{1k}^{(q)} & \ldots \\
\ldots
\end{array}
\right)
\label{eq.Gammachop}
\end{eqnarray}
would have been discarded. So it is the product of these bottom rows
of the $\Gamma$ matrix by the $K_k^{(q)}$ of higher order $k$, to be multiplied
by coefficients $\beta_k^q$, that has been ignored in the analysis of
Section \ref{sec.tayl}\@.
To re-insert their time-dependent randomness into the simulation,
one can build the dot product of the missing rows and columns in (\ref{eq.Gammachop}),
and multiply this vector with coefficients $\beta$ that are randomly
generated at each new image on the fly, consistent with the fact that
they have unknown history, that they are not carried over between the time steps
of the simulation, and that they have no associated eigenfrequencies in the
gradient matrix. The elements of the components of $\Gamma(D/\bar D)$ depend
on the ratio $D/\bar D$ and scale these random contributions appropriately
depending on how large the size $\bar D$ of the super-aperture must be
chosen to comply with (\ref{eq.Dbaroft}) during the next time interval.

In conclusion, feeding a stream of random kinks into the EOM of the
expansion coefficients that otherwise move on circles does not introduce
more parameters; it can be derived from a model of coupling the information
just outside the pupil into the actual diameter once the effect of
radial scaling on the basis functions has been calculated.

\section{Summary} 
The numerical synthesis of---in the Markov sense---independent phase
screens by superposition of statistically independent basis functions
over a telescope entrance pupil multiplied by a vector of Gaussian
random numbers is a known concept. Connection of these still pictures
to a movie needs some specification of the equation-of-motion of the
random numbers, which I have bound to the Taylor ``frozen'' screen
assumption that on short time scales an instance of a phase screen
is merely shifted.

If one external parameter---the wind velocity---is introduced, this
suffices to formulate a first-order differential equation in time
for the deterministic (``ballistic'') motion. A decoupling of
temporal and spatial structure is found by diagonalizing
the skew-symmetric matrix of the gradient operator. This
diagonalization needs to be done only once for each size of
the basis set (each Taylor movie), and
introduces new basis functions which undulate across the
pupil in the wind direction on scales determined by the eigenvalues
of the diagonalization.

Randomness in this scheme of oscillating expansion
coefficients (in the diagonalized coordinate system) is re-introduced
by virtual embedding of the pupil into a larger pupil which
regulates at which time scales and at which strength the
fluctuations outside the pupil---hence unknown---mix 
higher-order fluctuations into the time-dependent expansion coefficients.

The benefit of this technique is in the reduction of the problem of
two-dimensional time-dependent phase screens to the time-dependence of a vector
of a few expansion coefficients. This aims at more efficient implementation on a
computer and is equivalent to pulling a much larger sky linearly
across the circular telescope's aperture.
No new physics is introduced; on the contrary, the technique is largely
independent of modal decomposition and parametrization of the phase
structure function.

\begin{acknowledgments}
This work is supported by the NWO VICI grant
639.043.201
to A. Quirrenbach,
``Optical Interferometry: A new Method for Studies of Extrasolar Planets.''
\end{acknowledgments}

\appendix

\section{KL Integral Kernel} 

\subsection{Recursion with respect to azimuthal quantum number} \label{app.hyper} 

The kernel of (\ref{eq.kh}) and (\ref{eq.khsymm}),
\begin{equation}
-c_\varphi \frac{1}{2\openone_x}\int_0^{\openone_x}
\sqrt{xx'}S_q(x,x')
K_p^{(q)}(x')
dx'
=\lambda_{p,q}^2x K_{p}^{(q)}(x)
\label{eq.kh2}
,
\end{equation}
contains functions $S_q$ of the form
\begin{equation}
S_q(x,x')=
\frac{1}{(2\openone_x)^{2+\gamma}}
\int_0^{2\pi}d\theta \sqrt{xx'}(x^2+x'^2-2xx'\cos \theta)^{(1+\gamma)/2} \cos q\theta.
\label{eq.Sdef}
\end{equation}
Auxiliary polar coordinates
\begin{equation}
x\equiv 2\openone_x u\cos\alpha,\quad x'\equiv 2\openone_x u\sin\alpha,\quad \sin(2\alpha)\equiv z,
\label{eq.alpha}
\end{equation}
render this into hypergeometric functions
\begin{equation}
S_q
=
u^{2+\gamma}
\sqrt{\frac{z}{2}}
\int_0^{2\pi}d\theta (1-z\cos \theta)^{(1+\gamma)/2} \cos q\theta
=
u^{2+\gamma}
\sqrt{\frac{z}{2}}
2\pi \frac{(-1/2-\gamma/2)_q}{q!}
\left(\frac{z}{2}\right)^q
\,
_2F_1\left(
\begin{array}{c}
\frac{q}{2}-\frac{1+\gamma}{4}, \frac{q}{2}+\frac{1-\gamma}{4} \\
q+1
\end{array}
|z^2 \right)
\label{eq.Xof2f1}
\end{equation}
in the argument range $0\le z\le 1$.
The quadratic transformations of \cite[15.3.19--21]{AS}
are applicable or a re-writing as Legendre functions of \cite[15.4.10--11]{AS},
which may be used to tune into whatever numerical library is available.

\subsection{Fourier Transform}
The Fourier representation of the KL equation (\ref{eq.kh2}) is obtained by
switching the $(x,x')$ coordinates in $S_q$ to the
wave number domain introduced in Section \ref{sec.Kofsig}\@.
We summarize the transformation for $q\neq 0$.
First, the quadratic transform \cite[15.3.20]{AS} is applied to the hypergeometric function
(\ref{eq.Xof2f1})
\begin{equation}
S_q
=
u^{2+\gamma}
\sqrt{\frac{z}{2}}
2\pi \frac{(-1/2-\gamma/2)_q}{q!}
\left(\frac{z}{2}\right)^q
(1+z)^{-q+(1+\gamma)/2}
\,
_2F_1\left(
\begin{array}{c}
q-1/2-\gamma/2, q+1/2 \\
2q+1
\end{array}
|\frac{2z}{1+z} \right)
.
\label{eq.Sofz}
\end{equation}
Back to the $\alpha$-coordinates with
$z=2\sin\alpha\cos\alpha$ and $1+z=(\sin\alpha+\cos\alpha)^2$,
\begin{equation}
S_q
=
u^{2+\gamma}
(\sin\alpha\cos\alpha)^{q+1/2}
2\pi \frac{(-1/2-\gamma/2)_q}{q!}
(1+z)^{-q+(1+\gamma)/2}
\,
_2F_1\left(
\begin{array}{c}
q-1/2-\gamma/2, q+1/2 \\
2q+1
\end{array}
|\frac{4\sin\alpha\cos\alpha}{(\sin\alpha+\cos\alpha)^2} \right)
,
\end{equation}
its argument now matches the integral
\cite[6.576.2]{GR} provided $q>(1+\gamma)/2$
\begin{eqnarray}
&&
\int_0^\infty k^{-2-\gamma} J_q(\sin\alpha k)J_q(\cos\alpha k)dk
\nonumber
\\
&&\quad =
(\cos\alpha\sin\alpha)^q\frac{\Gamma(q-1/2-\gamma/2)}{2^{
2+\gamma}(\sin\alpha+\cos\alpha)^{2q-1-\gamma}q!\Gamma(3/2+\gamma/2)}
\,
_2F_1\left(
\begin{array}{c}
q-1/2-\gamma/2, q+1/2 \\
2q+1
\end{array}
|\frac{4\sin\alpha\cos\alpha}{(\sin\alpha+\cos\alpha)^2} \right)
.
\end{eqnarray}
The equivalent write-up
\begin{eqnarray}
&&
2^{2+\gamma}
\frac{\Gamma(3/2+\gamma/2)}{\Gamma(-1/2-\gamma/2)}
\int_0^\infty k^{-2-\gamma} J_q(\sin\alpha k)J_q(\cos\alpha k)dk
\nonumber
\\
&&\quad =
\left(\frac{z}{2}\right)^q\frac{(-1/2-\gamma/2)_q (1+z)^{-q+(1+\gamma)/2}}
{
q!
}
\,
_2F_1\left(
\begin{array}{c}
q-1/2-\gamma/2, q+1/2 \\
2q+1
\end{array}
|\frac{4\sin\alpha\cos\alpha}{(\sin\alpha+\cos\alpha)^2} \right)
\end{eqnarray}
allows replacement of the hypergeometric function by the $k$-integral in (\ref{eq.Sofz}),
\begin{equation}
S_q
=
u^{2+\gamma}(\sin\alpha\cos\alpha)^{1/2}2\pi
\,
2^{2+\gamma}
\frac{\Gamma(3/2+\gamma/2)}{\Gamma(-1/2-\gamma/2)}
\int_0^\infty k^{-2-\gamma} J_q(\sin\alpha k)J_q(\cos\alpha k)dk
.
\end{equation}
We substitute $(\sin\alpha\cos\alpha)^{1/2}=(xx')^{1/2}/(2\openone_x u)$
and insert $S_q$ back into (\ref{eq.kh2}),
\begin{equation}
-c_\varphi \frac{1}{(2\openone_x)^2}\int_0^{\openone_x}dx'
xx'
u^{1+\gamma}\pi
2^{3+\gamma}
\frac{\Gamma(3/2+\gamma/2)}{\Gamma(-1/2-\gamma/2)}
K_p^{(q)}(x')
\int_0^\infty k^{-2-\gamma} J_q(\sin\alpha k)J_q(\cos\alpha k)dk
=\lambda_{p,q}^2x K_{p}^{(q)}(x)
.
\end{equation}
The linear substitution $k=4\pi \openone_x u\sigma$ of the integration variable turns this into
\begin{equation}
-c_\varphi \frac{1}{(2\openone_x)^{3+\gamma}}\int_0^{\openone_x}dx'
xx'
\pi
2^2
\frac{\Gamma(3/2+\gamma/2)}{\Gamma(-1-\gamma/2)}
K_p^{(q)}(x')
\pi^{-1-\gamma}
\int_0^\infty \sigma^{-2-\gamma} J_q(2\pi x\sigma)J_q(2\pi x'\sigma)d\sigma
=\lambda_{p,q}^2x K_{p}^{(q)}(x)
.
\label{eq.mixdEV}
\end{equation}
The momentum representation of the KL equation is obtained by
multiplication with $J_q(2\pi\sigma'x)$ on both sides and integration 
over $x$ with (\ref{eq.Kofsigma}),
\begin{equation}
\frac{\pi}{\openone_x^{3+\gamma}}
\frac{(-c_\varphi)}{(2\pi)^{1+\gamma}}
\frac{\Gamma(3/2+\gamma/2)}{\Gamma(-1/2-\gamma/2)}
\int_0^{\openone_x}x\, dx
\int_0^\infty
d\sigma
\sigma^{-2-\gamma} J_q(2\pi x\sigma')J_q(2\pi x\sigma)
K_p^{(q)}(\sigma)
=\lambda_{p,q}^2 K_{p}^{(q)}(\sigma')
,
\end{equation}
where
\begin{equation}
-c_\varphi \frac{\Gamma(3/2+\gamma/2)}{(2\pi)^{1+\gamma} \Gamma(-1/2-\gamma/2)}
\approx 0.022656
\label{eq.const002}
\end{equation}
is a well-known factor \cite[(3)]{RoddierOE29}\cite[(25)]{NollJOSA66}
at $\gamma=2/3$\@.

\subsection{Connection to Noll's Covariances}
If the radial functions are expanded in an orthonormal Zernike basis \cite{NollJOSA66,MatharArxiv0705a},
\begin{equation}
\int_0^{\openone_x} x R_n^q(x/\openone_x)R_{n'}^q(x/\openone_x) dx=\frac{\delta_{n,n'}\openone_x^2}{2(n+1)};
\label{eq.Rortho}
\end{equation}
\begin{equation}
K_p^{(q)}(x) \equiv \sum_{n\equiv q(\mathrm{mod} 2)} \tau_{n,p,q} \frac{1}{\openone_x}\sqrt{2(n+1)} R_n^{q}(x/\openone_x),
\label{eq.KZexp}
\end{equation}
(\ref{eq.mixdEV}) becomes
\begin{eqnarray}
-c_\varphi \frac{\pi^{-\gamma}}{2^{1+\gamma}\openone_x^{3+\gamma}}
\frac{\Gamma(3/2+\gamma/2)}{\Gamma(-1/2-\gamma/2)}
\int_0^{\openone_x}dx'
xx'
\sum_n \tau_{n,p,q} \sqrt{2(n+1)}
R_n^{q}(x'/\openone_x)
\int_0^\infty \sigma^{-2-\gamma} J_q(2\pi x\sigma)J_q(2\pi x'\sigma)d\sigma
&&
\nonumber \\
=\lambda_{p,q}^2x
\sum_{n'} \tau_{n',p,q} \sqrt{2(n'+1)}
R_{n'}^{q}(x/\openone_x)
.
&&
\end{eqnarray}
Integration over $x'$ on the l.h.s.\ with
the Fourier representation (\ref{eq.Kofsigma})
\cite{NollJOSA66,Born,DaiOL31,CerjanJOSAA24}
\begin{equation}
\sum_n \tau_{n,p,q} 2\pi i^q\sqrt{2(n+1)} \int_0^{\openone_x} x R_n^{q}(x/\openone_x) J_q(2\pi\sigma x)dx
= 
\sum_n \tau_{n,p,q} i^n \openone_x^2 \sqrt{2(n+1)} \frac{J_{n+1}(2\pi\sigma\openone_x)}{\sigma\openone_x}
\label{eq.RFT}
\end{equation}
yields
\begin{eqnarray}
-c_\varphi \frac{\pi^{-\gamma}}{2^{1+\gamma}\openone_x^{3+\gamma}}
\frac{\Gamma(3/2+\gamma/2)}{\Gamma(-1/2-\gamma/2)}
x
\sum_n \tau_{n,p,q} \sqrt{2(n+1)}
i^{n-q} \openone_x
\int_0^\infty
\frac{J_{n+1}(2\pi\sigma\openone_x)}{2\pi\sigma}
\sigma^{-2-\gamma} J_q(2\pi x\sigma)d\sigma
\nonumber
&&
\\
=\lambda_{p,q}^2x
\sum_{n'} \tau_{n',p,q} \sqrt{2(n'+1)}
R_{n'}^{q}(x/\openone_x)
.
&&
\end{eqnarray}
Multiplication by $R_{n''}^q(x/\openone_x)$ on both sides and integration over $x$
employing
(\ref{eq.Rortho})
on the r.h.s.\ and integration over $x$ with (\ref{eq.RFT}) on the l.h.s.\
yields
\begin{eqnarray}
-c_\varphi \frac{\pi^{-\gamma}}{2^{1+\gamma}\openone_x^{3+\gamma}}
\frac{\Gamma(3/2+\gamma/2)}{\Gamma(-1/2-\gamma/2)}
\sum_n \tau_{n,p,q} \sqrt{2(n+1)}
i^{n-q} \openone_x
\int_0^\infty d\sigma
\frac{J_{n+1}(2\pi\sigma\openone_x)}{2\pi\sigma}
\sigma^{-2-\gamma}
i^{n''-q}\openone_x\frac{J_{n''+1}(2\pi\sigma\openone_x)}{2\pi\sigma}
\nonumber
&&
\\
=\lambda_{p,q}^2
\tau_{n'',p,q} \frac{\openone_x^2}{\sqrt{2(n''+1)}}
.
&&
\end{eqnarray}
The substitution $2\pi\sigma\openone_x=k$ of the integration variable generates
\begin{equation}
-c_\varphi \pi
\frac{\Gamma(3/2+\gamma/2)}{\Gamma(-1/2-\gamma/2)}
\sum_n \tau_{n,p,q} \sqrt{2(n+1)}
i^{n-q}
\int_0^\infty dk
J_{n+1}(k)
k^{-4-\gamma}
i^{n''-q}J_{n''+1}(k)
=\lambda_{p,q}^2
\tau_{n'',p,q} \frac{1}{\sqrt{2(n''+1)}}
.
\end{equation}
For $n+n'>1+\gamma>0$, the integral is \cite[6.574.2]{GR}
\begin{equation}
-c_\varphi 2\pi
\frac{\Gamma(3/2+\gamma/2)}{\Gamma(-1/2-\gamma/2)}
\sum_n \tau_{n,p,q} \sqrt{(n+1)(n'+1)}
(-1)^{(n+n')/2-q}
I_{nn'}
=\lambda_{p,q}^2
\tau_{n',p,q}
,
\label{eq.FriedZ}
\end{equation}
where \cite{NollJOSA66,BoremanJOSAA13}
\begin{equation}
I_{nn'}\equiv
\frac{\Gamma(4+\gamma)\Gamma\left(\frac{n+n'-1-\gamma}{2}\right)}
{2^{4+\gamma}
\Gamma\left(2+\frac{n'-n+1+\gamma}{2}\right)
\Gamma\left(3+\frac{n+n'+1+\gamma}{2}\right)
\Gamma\left(2+\frac{n-n'+1+\gamma}{2}\right)}
.
\end{equation}

This is consistent with Noll's equation (25), the Kolmogorov covariance
of a field with unit amplitude in terms of an orthogonal Zernike basis $\{Z_j\}$,
\begin{equation}
<a_j^*a_j'>
=\frac{2\times 0.023}{\pi}
\sqrt{(n+1)(n'+1)}(-1)^{(n+n')/2-q}
\left(\frac{D}{2r_0}\right)^{1+\gamma}
\underbrace{
\int d\sigma \sigma^{-2-\gamma}\frac{J_{n+1}(2\pi\sigma)J_{n'+1}(2\pi\sigma)}{\sigma^2}}_{(2\pi)^{3+\gamma}I_{nn'}},
\label{eq.Nollaa}
\end{equation}
re-interpreted with (\ref{eq.const002}),
\begin{equation}
<a_j^*a_j'>
=
-8c_\varphi \frac{\Gamma(3/2+\gamma/2)}{\Gamma(-1/2-\gamma/2)}
\sqrt{(n+1)(n'+1)}(-1)^{(n+n')/2-q}I_{nn'}\left(\frac{D}{r_0}\right)^{1+\gamma}
.
\end{equation}
This diverges from (\ref{eq.FriedZ}) by a nominal factor $4/\pi$,
\begin{itemize}
\item
the factor $4$ means that the $\tau_{n,p,q}$ are amplitudes to be post-multiplied
by $D^2$, Fried's equation (23), whereas (\ref{eq.Nollaa}) contains an implicit factor
of the squared radius which was silently dropped between Noll's (22) and (25).
\item
the unit amplitude in Noll's Zernike basis---which is normalized with
an additional weight $1/\pi$ in his (3) but not in my (\ref{eq.ortho})---
is equivalent to an amplitude $\sqrt \pi$ here, and introduces a factor $\pi$
because the covariance is a quadratic form of the amplitudes.
\end{itemize}

On the same basis, Dai's variable $c_0$ \cite[(2.10)]{DaiJOSAA12}
is consistent with the present
calculation. Roddier's expression $E(a_j,a_{j'})$
contains an additional factor $2^{1+\gamma}/\pi\approx 1.0106$,
and the same small factor is present
in Conan's version of (\ref{eq.const002}) \cite[(3.20)]{Conan00}\cite[(15)]{ConanJOSAA25}.
I am unable to trace these back---if applied it triggers a systematic increase
of all eigenvalues $\lambda_{p,q}^2$ by one percent.

\subsection{Piston Term at $q=0$} 

The case of the modes $K_p^{(0)}$ in (\ref{eq.kh}) induces three special terms proportional
to ${\cal G}_1$ and ${\cal G}_2$ after insertion of (\ref{eq.R0}).
Modes on both sides of the eigenvalue equation are expanded
in Zernike bases with (\ref{eq.KZexp}).

Isolation of a single $\tau$ on the right hand side, ie,  a matrix equation for
the $\tau$, is achieved by multiplication with $x K_{p'}^{(q)}(x)$ and integration
over $dx$.
A special case of (\ref{eq.Rortho}), namely
\[
\int_0^{\openone_x} x dx R_n^0(x/\openone_x)
=
\int_0^{\openone_x} x dx R_n^0(x/\openone_x)R_0^0(x/\openone_x)
=
\frac{\delta_{n,0}\openone_x^2}{2},
\]
reduces some integrals at the ${\cal G}_1$ and ${\cal G}_2$ to yield
\begin{eqnarray}
&&
-c_\varphi
8\pi
\sum_{n'\equiv 0(\mathrm{mod} 2)} \tau_{p,0,n'} \sqrt{n'+1}
\int u^{4+\gamma} du\,d\alpha\,
z
R_{n}^0(2u\cos\alpha)
R_{n'}^{0}(2u\sin\alpha)
\,_2F_1\left(\begin{array}{c}-1/4-\gamma/4,1/4-\gamma/4\\1\end{array}|z^2\right)
\nonumber \\
&&
+
\delta_{n,0}
c_\varphi 8\pi
\sum_{n'\equiv 0(\mathrm{mod} 2)} \tau_{p,0,n'} \sqrt{n'+1}
\int u^{4+\gamma} du\,d\alpha\,
z
R_{n'}^{0}(2u\cos\alpha)
\,_2F_1\left(\begin{array}{c}-1/4-\gamma/4,1/4-\gamma/4\\1\end{array}|z^2\right)
\nonumber \\
&&
+
\delta_{n',0} \tau_{p,0,0}
c_\varphi 8\pi
\int u^{4+\gamma} du\,d\alpha\,
z
R_{n}^0(2u\cos\alpha)
\,_2F_1\left(\begin{array}{c}-1/4-\gamma/4,1/4-\gamma/4\\1\end{array}|z^2\right)
\nonumber \\
&&
-
\frac{\pi}{4} {\cal G}_2
\delta_{n,0}
\delta_{n',0}
\tau_{p,0,0}
=
\lambda_{p,0}^2 
\tau_{p,0,n} \frac{1}{\sqrt{n+1}}
.
\label{eq.q0reduc}
\end{eqnarray}

The value of (\ref{eq.G2}) is a constant,
the residual of Dai's first mode \cite[Table I]{DaiJOSAA12}:
\begin{equation}
{\cal G}_2=c_\varphi 8
\frac{1}{\pi/4}
\,
\frac{1}{(2\openone_x)^{5+\gamma}}
\int_0^{\openone_x}dx x
\int_0^{\openone_x}dx'' x''
\int_0^{2\pi} d\theta
(x^2+x''^2-2xx''\cos\theta)^{(1+\gamma)/2}
\end{equation}
which is transformed to the polar coordinates proposed in (\ref{eq.alpha}),
with Jacobian $(2\openone_x)^2 u$,
\begin{equation}
\frac{1}{(2\openone_x)^2}\int_0^{\openone_x}dx\int_0^{\openone_x}dx'' = 
\int_0^{\pi/4}d\alpha \int_0^{1/(2\cos\alpha)}udu
+
\int_{\pi/4}^{\pi/2}d\alpha \int_0^{1/(2\sin\alpha)}udu.
\end{equation}
\begin{eqnarray}
{\cal G}_2
&=&
c_\varphi \frac{16}{\pi}
\int_{\begin{array}{c}0\le u\sin\alpha\le 1/2 \\ 0\le u\cos\alpha\le 1/2\end{array}}
 udu \int_0^{\pi/2} d\alpha u^2\sin(2\alpha)
u^{1+\gamma}\int_0^{2\pi}d\theta [1-\sin(2\alpha)\cos\theta]^{(1+\gamma)/2} \\
&=&
\frac{16}{\pi}c_\varphi
\int u^{4+\gamma}du \int d\alpha z
(2\pi)\,_2F_1\left(\begin{array}{c}-1/4-\gamma/4,-\gamma/4+1/4\\1\end{array}|z^2\right) \\
&=&
64 c_\varphi
\int_0^{\pi/4} d\alpha z
\frac{1}{(5+\gamma)(2\cos\alpha)^{5+\gamma}}
\,_2F_1\left(\begin{array}{c}-1/4-\gamma/4,1/4-\gamma/4\\1\end{array}|z^2\right)
\\
&=&
4 c_\varphi
\frac{1}{(5+\gamma)2^{1+\gamma}}\,\frac{\Gamma(3+\gamma)}{\Gamma^2(5/2+\gamma/2)}
=
8c_\varphi\frac{\Gamma(3/2+\gamma/2)}{\Gamma(-1/2-\gamma/2)}I_{00}
\approx
0.299953532617054
c_\varphi
\approx
1.032421639
.
\end{eqnarray}
Since $R_0^0=1$, within the top row $n=0$ of the coefficients matrix,  the first
two and the last two lines of (\ref{eq.q0reduc}) cancel. Within the left column $n'=0$ of the
coefficients matrix, the third line cancels the first and the fourth line cancels the second.
This is the real-space equivalent to Noll's removal of the piston mode in Fourier
space \cite{NollJOSA66}, which eliminates the $R_0^0$ term
from the $q=0$ block of the KL equation.

\section{Gradient Matrix in Wave number Space}\label{sec.OmegaFT}

The Fourier integral of the EOM (\ref{eq.ddtphi}) transforms the gradient into
a multiplication with the wave number $-2\pi i\bm{\sigma}$,
\begin{equation}
\sum_{l,m}\frac{d\beta_l^{(m)}(t)}{dt}\lambda_{l,m}K_l^{(m)}(\bm{\sigma})
=-\bm{v}\cdot
(-2\pi i)\bm{\sigma}
\frac{2\openone_x}{D}
\sum_{j,n}\beta_j^{(n)}(t)\lambda_{j,n}K_j^{(n)}(\bm{\sigma}).
\end{equation}
As before we assume the wind velocity ${\bm v}$ contains only a component
into the $X$ direction,
\begin{equation}
\bm{v}\cdot \bm{\sigma}
= v\sigma \cos\theta_\sigma,
\end{equation}
and (\ref{eq.Fofsigma}) yields
\begin{equation}
\sum_{l,m}\frac{d\beta_l^{(m)}(t)}{dt}\lambda_{l,m}K_l^{(m)}(\sigma)M_m(\theta_\sigma)
=2\pi i v
\frac{2\openone_x}{D}
\sigma \cos\theta_\sigma
\sum_{j,n}\beta_j^{(n)}(t)\lambda_{j,n}K_j^{(n)}(\sigma)M_n(\theta_\sigma).
\end{equation}
After insertion of (\ref{eq.Mtheta}),
\begin{equation}
\sum_{l,m}\frac{d\beta_l^{(m)}(t)}{dt}\lambda_{l,m}K_l^{(m)}(\sigma)
\sqrt{\frac{\epsilon_m}{2\pi}}
\left\{
\begin{array}{c}
\cos m\theta_\sigma\\
\sin m\theta_\sigma\\
\end{array}
\right.
=\pi i v
\frac{2\openone_x}{D}
\sigma
\sum_{j,n}\beta_j^{(n)}(t)
\lambda_{j,n} K_j^{(n)}(\sigma)
\sqrt{\frac{\epsilon_n}{2\pi}}
\left\{
\begin{array}{c}
\cos(n+1)\theta_\sigma + \cos(n-1)\theta_\sigma \\
\sin(n+1)\theta_\sigma + \sin(n-1)\theta_\sigma \\
\end{array}
\right.
\end{equation}
we project this onto a single component of the $\beta$-vector by
multiplication
with $\sigma \lambda_{k,s}K_k^{(s)*}(\sigma)M_k(\theta_\sigma)$ and integration
over $d^2\sigma$ using the orthogonality (\ref{eq.KorthoF}) on the left hand side,
\begin{equation}
\frac{D}{2\openone_x \pi iv}
\frac{d\beta_k^{(s)}(t)}{dt}\lambda_{k,s}^2
=
\sum_{j,n}\beta_j^{(n)}(t)
\frac{2}{\sqrt{\epsilon_n\epsilon_s}}(\delta_{n+1,s}+\delta_{n-1,s})(1-\delta_{s+n,-1})
\int_0^\infty d\sigma
\sigma^2
\lambda_{j,n} K_j^{(n)}(\sigma)
\lambda_{k,s} K_k^{(s)*}(\sigma)
\label{eq.dipolLadd}
.
\end{equation}
For a Zernike decomposition (\ref{eq.RFT}),
the integrals $\int_0^\infty J_{j'+1}(\sigma)J_{k'+1}(\sigma)d\sigma$
can be evaluated analytically \cite[p. 50]{MO2Afl},
\begin{eqnarray}
\frac{D}{2 v}
\frac{d\beta_k^{(s)}(t)}{dt}\lambda_{k,s}^2
& =&
\sum_{j,n}\beta_j^{(n)}(t)
\frac{1}{\sqrt{\epsilon_n\epsilon_s}}(\delta_{n+1,s}+\delta_{n-1,s})(1-\delta_{s+n,-1})
\lambda_{j,n}
\lambda_{k,s}
\nonumber
\\
&&\times
\sum_{j'k'}
(-1)^{(j'-k'+|j'-k'|)/2}
\tau_{j',j,n} \sqrt{(j'+1)(k'+1)}
\tau_{k',k,s},
\end{eqnarray}
which recovers the elements of Noll's gradient matrix $\gamma_{jj'}^x$ \cite{NollJOSA66}.

\section{Zernike Covariances} \label{zern.app}

A table of the dominant KL eigenvectors $K_p^{(q)}M_q$
is attached,
diagonalizing a
$80\times 80$
matrix for each individual $|q|$.
It contains the eigenvalue, after a colon the value of $|q|$, then the
expansion in Zernike polynomials in Noll's nomenclature. For $q=0$, the
value in parentheses is the index of $Z$, for $q\neq 0$ a comma-separated
pair of first the index for the cosine-term, $q>0$, then the index
for the sine-term, $q<0$.

Terms with expansion coefficients
down to a threshold of $10^{-6}$ are listed.
The squared expansion coefficients sum to
$1/\pi$
which implies that the
square of the eigenvector integrated over the pupil is normalized
to
unity.

\small
\begin{verbatim}
0.3529054 :  1:  +0.5639006*Z(2,3) -0.0180275*Z(8,7) +0.0010116*Z(16,17) +0.0000047*Z(30,29) 
+0.0000011*Z(46,47)

0.0187926 :  2:  +0.5550339*Z(6,5) -0.1006468*Z(12,13) +0.0108338*Z(24,23) -0.0003738*Z(38,39) 
+0.0000221*Z(58,57) +0.0000013*Z(80,81)

0.0187926 :  0:  +0.5550339*Z(1) -0.1006468*Z(4) +0.0108338*Z(11) -0.0003738*Z(22) +0.0000221*Z(37) 
+0.0000013*Z(56)

0.0052209 :  3:  +0.5415715*Z(10,9) -0.1562750*Z(18,19) +0.0241979*Z(32,31) -0.0016589*Z(48,49) 
+0.0001053*Z(70,69)

0.0048842 :  1:  +0.0175046*Z(2,3) +0.5382264*Z(8,7) -0.1661406*Z(16,17) +0.0266343*Z(30,29) 
-0.0019283*Z(46,47) +0.0001237*Z(68,67)

0.0021547 :  4:  +0.5274038*Z(14,15) -0.1965600*Z(26,25) +0.0387868*Z(40,41) -0.0038510*Z(60,59) 
+0.0003027*Z(82,83) -0.0000093*Z(110,109) +0.0000018*Z(140,141)

0.0016529 :  2:  +0.0944936*Z(6,5) +0.4941255*Z(12,13) -0.2485822*Z(24,23) +0.0581153*Z(38,39) 
-0.0070710*Z(58,57) +0.0006213*Z(80,81) -0.0000294*Z(108,107) +0.0000031*Z(138,139)

0.0016529 :  0:  +0.0944936*Z(1) +0.4941255*Z(4) -0.2485822*Z(11) +0.0581153*Z(22) -0.0070710*Z(37) 
+0.0006213*Z(56) -0.0000294*Z(79) +0.0000031*Z(106)

0.0010850 :  5:  +0.5136656*Z(20,21) -0.2270441*Z(34,33) +0.0534941*Z(50,51) -0.0068153*Z(72,71) 
+0.0006519*Z(96,97) -0.0000346*Z(126,125) +0.0000037*Z(158,159)

0.0007763 :  3:  +0.1423413*Z(10,9) +0.4455925*Z(18,19) -0.3017149*Z(32,31) +0.0907617*Z(48,49) 
-0.0149565*Z(70,69) +0.0017093*Z(94,95) -0.0001269*Z(124,123) +0.0000101*Z(156,157)

0.0007595 :  1:  +0.0040371*Z(2,3) +0.1508302*Z(8,7) +0.4388195*Z(16,17) -0.3064797*Z(30,29) 
+0.0938139*Z(46,47) -0.0157202*Z(68,67) +0.0018198*Z(92,93) -0.0001376*Z(122,121) +0.0000109*Z(154,155)

0.0006175 :  6:  +0.5007017*Z(28,27) -0.2508042*Z(42,43) +0.0677902*Z(62,61) -0.0103942*Z(84,85) 
+0.0011738*Z(112,111) -0.0000838*Z(142,143) +0.0000076*Z(178,177)

0.0004293 :  4:  +0.1743752*Z(14,15) +0.3990822*Z(26,25) -0.3363293*Z(40,41) +0.1220098*Z(60,59) 
-0.0249092*Z(82,83) +0.0034953*Z(110,109) -0.0003387*Z(140,141) +0.0000288*Z(176,175)

0.0003857 :  2:  -0.0317365*Z(6,5) -0.2140331*Z(12,13) -0.3498546*Z(24,23) +0.3572055*Z(38,39) 
-0.1430459*Z(58,57) +0.0318299*Z(80,81) -0.0047925*Z(108,107) +0.0005028*Z(138,139) -0.0000441*Z(174,173) 
+0.0000017*Z(212,213)

0.0003857 :  0:  -0.0317365*Z(1) -0.2140331*Z(4) -0.3498546*Z(11) +0.3572055*Z(22) -0.1430459*Z(37) 
+0.0318299*Z(56) -0.0047925*Z(79) +0.0005028*Z(106) -0.0000441*Z(137) +0.0000017*Z(172)

0.0003822 :  7:  +0.4885875*Z(36,35) -0.2697207*Z(52,53) +0.0814295*Z(74,73) -0.0144431*Z(98,99) 
+0.0018757*Z(128,127) -0.0001646*Z(160,161) +0.0000148*Z(198,197)

0.0002627 :  5:  -0.1968506*Z(20,21) -0.3562481*Z(34,33) +0.3585383*Z(50,51) -0.1508079*Z(72,71) 
+0.0363009*Z(96,97) -0.0060054*Z(126,125) +0.0007061*Z(158,159) -0.0000683*Z(196,195) +0.0000037*Z(236,237)

0.0002516 :  8:  +0.4773002*Z(44,45) -0.2850218*Z(64,63) +0.0943101*Z(86,87) -0.0188400*Z(114,113) 
+0.0027559*Z(144,145) -0.0002835*Z(180,179) +0.0000270*Z(218,219)

0.0002258 :  3:  -0.0579989*Z(10,9) -0.2465801*Z(18,19) -0.2674992*Z(32,31) +0.3804364*Z(48,49) 
-0.1874358*Z(70,69) +0.0513178*Z(94,95) -0.0094502*Z(124,123) +0.0012400*Z(156,157) -0.0001291*Z(194,193) 
+0.0000090*Z(234,235) -0.0000012*Z(280,279)

0.0002234 :  1:  -0.0015503*Z(2,3) -0.0625126*Z(8,7) -0.2490043*Z(16,17) -0.2604727*Z(30,29) 
+0.3814605*Z(46,47) -0.1902056*Z(68,67) +0.0525678*Z(92,93) -0.0097567*Z(122,121) +0.0012903*Z(154,155) 
-0.0001351*Z(192,191) +0.0000095*Z(232,233) -0.0000012*Z(278,277)

0.0001735 :  9:  +0.4667824*Z(54,55) -0.2975492*Z(76,75) +0.1064060*Z(100,101) -0.0234853*Z(130,129) 
+0.0038066*Z(162,163) -0.0004455*Z(200,199) +0.0000458*Z(240,241) -0.0000022*Z(286,285)

0.0001723 :  6:  -0.2131220*Z(28,27) -0.3173321*Z(42,43) +0.3721298*Z(62,61) -0.1768017*Z(84,85) 
+0.0486138*Z(112,111) -0.0092134*Z(142,143) +0.0012625*Z(178,177) -0.0001388*Z(216,217) +0.0000104*Z(260,259) 
-0.0000013*Z(306,307)

0.0001445 :  4:  -0.0803076*Z(14,15) -0.2618432*Z(26,25) -0.1948668*Z(40,41) +0.3859041*Z(60,59) 
-0.2255185*Z(82,83) +0.0727395*Z(110,109) -0.0157444*Z(140,141) +0.0024554*Z(176,175) -0.0002991*Z(214,215) 
+0.0000271*Z(258,257) -0.0000028*Z(304,305)

0.0001363 :  2:  -0.0144405*Z(6,5) -0.1076272*Z(12,13) -0.2659802*Z(24,23) -0.1497605*Z(38,39) 
+0.3840616*Z(58,57) -0.2422123*Z(80,81) +0.0825518*Z(108,107) -0.0187019*Z(138,139) +0.0030449*Z(174,173) 
-0.0003847*Z(212,213) +0.0000366*Z(256,255) -0.0000036*Z(302,303)

0.0001363 :  0:  -0.0144405*Z(1) -0.1076272*Z(4) -0.2659802*Z(11) -0.1497605*Z(22) +0.3840616*Z(37) 
-0.2422123*Z(56) +0.0825518*Z(79) -0.0187019*Z(106) +0.0030449*Z(137) -0.0003847*Z(172) +0.0000366*Z(211) 
-0.0000036*Z(254)

0.0001242 : 10:  +0.4569675*Z(66,65) -0.3079023*Z(88,89) +0.1177299*Z(116,115) -0.0282996*Z(146,147) 
+0.0050164*Z(182,181) -0.0006545*Z(220,221) +0.0000729*Z(264,263) -0.0000048*Z(310,311)

0.0001190 :  7:  -0.2251569*Z(36,35) -0.2821207*Z(52,53) +0.3795853*Z(74,73) -0.1999679*Z(98,99) 
+0.0614426*Z(128,127) -0.0130623*Z(160,161) +0.0020318*Z(198,197) -0.0002511*Z(238,239) +0.0000229*Z(284,283) 
-0.0000025*Z(332,333)

0.0000983 :  5:  -0.0988260*Z(20,21) -0.2668319*Z(34,33) -0.1319724*Z(50,51) +0.3797889*Z(72,71) 
-0.2570938*Z(96,97) +0.0949715*Z(126,125) -0.0235100*Z(158,159) +0.0042219*Z(196,195) -0.0005896*Z(236,237) 
+0.0000636*Z(282,281) -0.0000065*Z(330,331)

0.0000917 : 11:  +0.4477892*Z(78,77) -0.3165203*Z(102,103) +0.1283147*Z(132,131) -0.0332196*Z(164,165) 
+0.0063723*Z(202,201) -0.0009130*Z(242,243) +0.0001100*Z(288,287) -0.0000089*Z(336,337) +0.0000014*Z(390,389)

0.0000901 :  3:  -0.0293974*Z(10,9) -0.1411969*Z(18,19) -0.2594816*Z(32,31) -0.0576767*Z(48,49) 
+0.3648102*Z(70,69) -0.2824450*Z(94,95) +0.1141543*Z(124,123) -0.0303990*Z(156,157) +0.0058362*Z(194,193) 
-0.0008646*Z(234,235) +0.0000997*Z(280,279) -0.0000103*Z(328,329)

0.0000895 :  1:  -0.0007599*Z(2,3) -0.0319163*Z(8,7) -0.1439972*Z(16,17) -0.2582430*Z(30,29) 
-0.0521890*Z(46,47) +0.3632810*Z(68,67) -0.2842444*Z(92,93) +0.1156663*Z(122,121) -0.0309709*Z(154,155) 
+0.0059756*Z(192,191) -0.0008892*Z(232,233) +0.0001031*Z(278,277) -0.0000107*Z(326,327)

0.0000855 :  8:  -0.2341860*Z(44,45) -0.2502655*Z(64,63) +0.3825946*Z(86,87) -0.2204388*Z(114,113) 
+0.0744793*Z(144,145) -0.0174801*Z(180,179) +0.0030287*Z(218,219) -0.0004158*Z(262,261) +0.0000439*Z(308,309) 
-0.0000047*Z(360,359)

0.0000700 :  6:  -0.1141626*Z(28,27) -0.2655091*Z(42,43) -0.0779168*Z(62,61) +0.3661379*Z(84,85) 
-0.2825223*Z(112,111) +0.1171858*Z(142,143) -0.0325258*Z(178,177) +0.0065797*Z(216,217) -0.0010345*Z(260,259) 
+0.0001279*Z(306,307) -0.0000140*Z(358,357)

0.0000694 : 12:  +0.4391862*Z(90,91) -0.3237332*Z(118,117) +0.1382027*Z(148,149) -0.0381955*Z(184,183) 
+0.0078604*Z(222,223) -0.0012226*Z(266,265) +0.0001586*Z(312,313) -0.0000148*Z(364,363) +0.0000020*Z(418,419)

0.0000634 :  9:  -0.2410183*Z(54,55) -0.2213988*Z(76,75) +0.3823434*Z(100,101) -0.2384139*Z(130,129) 
+0.0874939*Z(162,163) -0.0223892*Z(200,199) +0.0042602*Z(240,241) -0.0006422*Z(286,285) +0.0000761*Z(334,335) 
-0.0000084*Z(388,387)

0.0000629 :  4:  -0.0439206*Z(14,15) -0.1651472*Z(26,25) -0.2407349*Z(40,41) +0.0166192*Z(60,59) 
+0.3327821*Z(82,83) -0.3113005*Z(110,109) +0.1454590*Z(140,141) -0.0443771*Z(176,175) +0.0097695*Z(214,215) 
-0.0016568*Z(258,257) +0.0002217*Z(304,305) -0.0000253*Z(356,355) +0.0000020*Z(410,411)

0.0000607 :  2:  +0.0077992*Z(6,5) +0.0613337*Z(12,13) +0.1781334*Z(24,23) +0.2255147*Z(38,39) 
-0.0478290*Z(58,57) -0.3169309*Z(80,81) +0.3197894*Z(108,107) -0.1560387*Z(138,139) +0.0492359*Z(174,173) 
-0.0111660*Z(212,213) +0.0019449*Z(256,255) -0.0002674*Z(302,303) +0.0000311*Z(354,353) -0.0000026*Z(408,409)

0.0000607 :  0:  +0.0077992*Z(1) +0.0613337*Z(4) +0.1781334*Z(11) +0.2255147*Z(22) -0.0478290*Z(37) 
-0.3169309*Z(56) +0.3197894*Z(79) -0.1560387*Z(106) +0.0492359*Z(137) -0.0111660*Z(172) +0.0019449*Z(211) 
-0.0002674*Z(254) +0.0000311*Z(301) -0.0000026*Z(352)

0.0000536 : 13:  +0.4311032*Z(104,105) -0.3297940*Z(134,133) +0.1474395*Z(166,167) -0.0431883*Z(204,203) 
+0.0094668*Z(244,245) -0.0015839*Z(290,289) +0.0002200*Z(338,339) -0.0000230*Z(392,391) +0.0000029*Z(448,449)

0.0000516 :  7:  -0.1269134*Z(36,35) -0.2602434*Z(52,53) -0.0315944*Z(74,73) +0.3476608*Z(98,99) 
-0.3023922*Z(128,127) +0.1387961*Z(160,161) -0.0425527*Z(198,197) +0.0095407*Z(238,239) -0.0016640*Z(284,283) 
+0.0002307*Z(332,333) -0.0000274*Z(386,385) +0.0000023*Z(442,443)

0.0000482 : 10:  -0.2462062*Z(66,65) -0.1951759*Z(88,89) +0.3796826*Z(116,115) -0.2541152*Z(146,147) 
+0.1003177*Z(182,181) -0.0277123*Z(220,221) +0.0057268*Z(264,263) -0.0009388*Z(310,311) +0.0001226*Z(362,361) 
-0.0000143*Z(416,417)

0.0000458 :  5:  +0.0572248*Z(20,21) +0.1816940*Z(34,33) +0.2158725*Z(50,51) -0.0753939*Z(72,71) 
-0.2938276*Z(96,97) +0.3299769*Z(126,125) -0.1751494*Z(158,159) +0.0600933*Z(196,195) -0.0148722*Z(236,237) 
+0.0028343*Z(282,281) -0.0004291*Z(330,331) +0.0000545*Z(384,383) -0.0000053*Z(440,441)

0.0000433 :  3:  +0.0170071*Z(10,9) +0.0872520*Z(18,19) +0.1958042*Z(32,31) +0.1813467*Z(48,49) 
-0.1207320*Z(70,69) -0.2590231*Z(94,95) +0.3385998*Z(124,123) -0.1938146*Z(156,157) +0.0703626*Z(194,193) 
-0.0182822*Z(234,235) +0.0036395*Z(280,279) -0.0005748*Z(328,329) +0.0000755*Z(382,381) -0.0000078*Z(438,439)

0.0000431 :  1:  +0.0004302*Z(2,3) +0.0185360*Z(8,7) +0.0894311*Z(16,17) +0.1964790*Z(30,29) 
+0.1784645*Z(46,47) -0.1238988*Z(68,67) -0.2561964*Z(92,93) +0.3390955*Z(122,121) -0.1952323*Z(154,155) 
+0.0711828*Z(192,191) -0.0185635*Z(232,233) +0.0037077*Z(278,277) -0.0005874*Z(326,327) +0.0000774*Z(380,379) 
-0.0000080*Z(436,437)

0.0000422 : 14:  +0.4234909*Z(120,119) -0.3349000*Z(150,151) +0.1560712*Z(186,185) -0.0481676*Z(224,225) 
+0.0111783*Z(268,267) -0.0019968*Z(314,315) +0.0002955*Z(366,365) -0.0000339*Z(420,421) +0.0000042*Z(480,479)

0.0000391 :  8:  -0.1375743*Z(44,45) -0.2525093*Z(64,63) +0.0080638*Z(86,87) +0.3262011*Z(114,113) 
-0.3173578*Z(144,145) +0.1594031*Z(180,179) -0.0533561*Z(218,219) +0.0130941*Z(262,261) -0.0025033*Z(308,309) 
+0.0003830*Z(360,359) -0.0000495*Z(414,415) +0.0000049*Z(474,473)

0.0000375 : 11:  -0.2501393*Z(78,77) -0.1712883*Z(102,103) +0.3752347*Z(132,131) -0.2677650*Z(164,165) 
+0.1128295*Z(202,201) -0.0333760*Z(242,243) +0.0074239*Z(288,287) -0.0013126*Z(336,337) +0.0001867*Z(390,389) 
-0.0000232*Z(446,447) +0.0000020*Z(508,507)

0.0000344 :  6:  +0.0691358*Z(28,27) +0.1926538*Z(42,43) +0.1883543*Z(62,61) -0.1211123*Z(84,85) 
-0.2517239*Z(112,111) +0.3399261*Z(142,143) -0.2023814*Z(178,177) +0.0770075*Z(216,217) -0.0211096*Z(260,259) 
+0.0044561*Z(306,307) -0.0007503*Z(358,357) +0.0001051*Z(412,413) -0.0000118*Z(472,471) +0.0000014*Z(534,535)

0.0000337 : 15:  +0.4163054*Z(136,135) -0.3392075*Z(168,169) +0.1641421*Z(206,205) -0.0531094*Z(246,247) 
+0.0129820*Z(292,291) -0.0024606*Z(340,341) +0.0003862*Z(394,393) -0.0000479*Z(450,451) +0.0000060*Z(512,511)

0.0000321 :  4:  +0.0267586*Z(14,15) +0.1087334*Z(26,25) +0.2017551*Z(40,41) +0.1340015*Z(60,59) 
-0.1709979*Z(82,83) -0.1972564*Z(110,109) +0.3432679*Z(140,141) -0.2272400*Z(176,175) +0.0933061*Z(214,215) 
-0.0272720*Z(258,257) +0.0060951*Z(304,305) -0.0010833*Z(356,355) +0.0001592*Z(410,411) -0.0000191*Z(470,469) 
+0.0000022*Z(532,533)

0.0000313 :  2:  +0.0047026*Z(6,5) +0.0382423*Z(12,13) +0.1210695*Z(24,23) +0.2006325*Z(38,39) 
+0.1125597*Z(58,57) -0.1854229*Z(80,81) -0.1762032*Z(108,107) +0.3424840*Z(138,139) -0.2359329*Z(174,173) 
+0.0996008*Z(212,213) -0.0297915*Z(256,255) +0.0067958*Z(302,303) -0.0012312*Z(354,353) +0.0001841*Z(408,409) 
-0.0000225*Z(468,467) +0.0000026*Z(530,531)

0.0000313 :  0:  +0.0047026*Z(1) +0.0382423*Z(4) +0.1210695*Z(11) +0.2006325*Z(22) +0.1125597*Z(37) 
-0.1854229*Z(56) -0.1762032*Z(79) +0.3424840*Z(106) -0.2359329*Z(137) +0.0996008*Z(172) -0.0297915*Z(211) 
+0.0067958*Z(254) -0.0012312*Z(301) +0.0001841*Z(352) -0.0000225*Z(407) +0.0000026*Z(466)

0.0000304 :  9:  +0.1465398*Z(54,55) +0.2432536*Z(76,75) -0.0420131*Z(100,101) -0.3030268*Z(130,129) 
+0.3280619*Z(162,163) -0.1787481*Z(200,199) +0.0647183*Z(240,241) -0.0172125*Z(286,285) +0.0035720*Z(334,335) 
-0.0005961*Z(388,387) +0.0000836*Z(444,445) -0.0000094*Z(506,505) +0.0000012*Z(570,571)

0.0000297 : 12:  -0.2530992*Z(90,91) -0.1494654*Z(118,117) +0.3694619*Z(148,149) -0.2795742*Z(184,183) 
+0.1249441*Z(222,223) -0.0393127*Z(266,265) +0.0093431*Z(312,313) -0.0017691*Z(364,363) +0.0002714*Z(418,419) 
-0.0000360*Z(478,477) +0.0000035*Z(540,541)

0.0000273 : 16:  +0.4095083*Z(152,153) -0.3428415*Z(188,187) +0.1716943*Z(226,227) -0.0579956*Z(270,269) 
+0.0148663*Z(316,317) -0.0029741*Z(368,367) +0.0004929*Z(422,423) -0.0000656*Z(482,481) +0.0000085*Z(544,545)

0.0000265 :  7:  +0.0797010*Z(36,35) +0.1994106*Z(52,53) +0.1601663*Z(74,73) -0.1560338*Z(98,99) 
-0.2088996*Z(128,127) +0.3425979*Z(160,161) -0.2266633*Z(198,197) +0.0946223*Z(238,239) -0.0284013*Z(284,283) 
+0.0065649*Z(332,333) -0.0012137*Z(386,385) +0.0001862*Z(442,443) -0.0000235*Z(504,503) +0.0000028*Z(568,569)

0.0000245 :  5:  +0.0363113*Z(20,21) +0.1259883*Z(34,33) +0.1996065*Z(50,51) +0.0877660*Z(72,71) 
-0.2031331*Z(96,97) -0.1363122*Z(126,125) +0.3365674*Z(158,159) -0.2553567*Z(196,195) +0.1170904*Z(236,237) 
-0.0379733*Z(282,281) +0.0093975*Z(330,331) -0.0018516*Z(384,383) +0.0003010*Z(440,441) -0.0000405*Z(502,501) 
+0.0000050*Z(566,567)

0.0000240 : 10:  +0.1541205*Z(66,65) +0.2330979*Z(88,89) -0.0710791*Z(116,115) -0.2790161*Z(146,147) 
+0.3351005*Z(182,181) -0.1966763*Z(220,221) +0.0764443*Z(264,263) -0.0218564*Z(310,311) +0.0048838*Z(362,361) 
-0.0008808*Z(416,417) +0.0001332*Z(476,475) -0.0000165*Z(538,539) +0.0000021*Z(606,605)

0.0000239 : 13:  -0.2552942*Z(104,105) -0.1294720*Z(134,133) +0.3627115*Z(166,167) -0.2897374*Z(204,203) 
+0.1366035*Z(244,245) -0.0454611*Z(290,289) +0.0114734*Z(338,339) -0.0023126*Z(392,391) +0.0003798*Z(448,449) 
-0.0000535*Z(510,509) +0.0000059*Z(574,575)

0.0000235 :  3:  +0.0107543*Z(10,9) +0.0575014*Z(18,19) +0.1434604*Z(32,31) +0.1897436*Z(48,49) 
+0.0490395*Z(70,69) -0.2179204*Z(94,95) -0.0977964*Z(124,123) +0.3285876*Z(156,157) -0.2685761*Z(194,193) 
+0.1292745*Z(234,235) -0.0436019*Z(280,279) +0.0111673*Z(328,329) -0.0022712*Z(382,381) +0.0003801*Z(438,439) 
-0.0000528*Z(500,499) +0.0000066*Z(564,565)

0.0000234 :  1:  +0.0002680*Z(2,3) +0.0117488*Z(8,7) +0.0591127*Z(16,17) +0.1445874*Z(30,29) 
+0.1887365*Z(46,47) +0.0460816*Z(68,67) -0.2187900*Z(92,93) -0.0948401*Z(122,121) +0.3278423*Z(154,155) 
-0.2695379*Z(192,191) +0.1302224*Z(232,233) -0.0440532*Z(278,277) +0.0113124*Z(326,327) -0.0023063*Z(380,379) 
+0.0003868*Z(436,437) -0.0000538*Z(498,497) +0.0000067*Z(562,563)

0.0000224 : 17:  +0.4030654*Z(170,171) -0.3459037*Z(208,207) +0.1787669*Z(248,249) -0.0628120*Z(294,293) 
+0.0168202*Z(342,343) -0.0035357*Z(396,395) +0.0006164*Z(452,453) -0.0000873*Z(514,513) +0.0000116*Z(578,579)

0.0000209 :  8:  +0.0890428*Z(44,45) +0.2029997*Z(64,63) +0.1324492*Z(86,87) -0.1821037*Z(114,113) 
-0.1668971*Z(144,145) +0.3393270*Z(180,179) -0.2477570*Z(218,219) +0.1125023*Z(262,261) -0.0366365*Z(308,309) 
+0.0091872*Z(360,359) -0.0018463*Z(414,415) +0.0003077*Z(474,473) -0.0000427*Z(536,537) +0.0000054*Z(604,603)

0.0000194 : 14:  -0.2568812*Z(120,119) -0.1111043*Z(150,151) +0.3552477*Z(186,185) -0.2984316*Z(224,225) 
+0.1477708*Z(268,267) -0.0517665*Z(314,315) +0.0138018*Z(366,365) -0.0029460*Z(420,421) +0.0005147*Z(480,479) 
-0.0000769*Z(542,543) +0.0000093*Z(610,609) -0.0000013*Z(680,681)

0.0000193 : 11:  +0.1605609*Z(78,77) +0.2224571*Z(102,103) -0.0959649*Z(132,131) -0.2547779*Z(164,165) 
+0.3390096*Z(202,201) -0.2131084*Z(242,243) +0.0883640*Z(288,287) -0.0269784*Z(336,337) +0.0064472*Z(390,389) 
-0.0012473*Z(446,447) +0.0002022*Z(508,507) -0.0000272*Z(572,573) +0.0000035*Z(642,641)

0.0000191 :  6:  +0.0453401*Z(28,27) +0.1395506*Z(42,43) +0.1919744*Z(62,61) +0.0447795*Z(84,85) 
-0.2210912*Z(112,111) -0.0789246*Z(142,143) +0.3210884*Z(178,177) -0.2777758*Z(216,217) +0.1408688*Z(260,259) 
-0.0501513*Z(306,307) +0.0135944*Z(358,357) -0.0029353*Z(412,413) +0.0005225*Z(472,471) -0.0000776*Z(534,535) 
+0.0000102*Z(602,601)

0.0000186 : 18:  +0.3969465*Z(190,189) -0.3484772*Z(228,229) +0.1853962*Z(272,271) -0.0675481*Z(318,319) 
+0.0188337*Z(370,369) -0.0041438*Z(424,425) +0.0007572*Z(484,483) -0.0001134*Z(546,547) +0.0000157*Z(614,613) 
-0.0000013*Z(684,685)

0.0000182 :  4:  +0.0175669*Z(14,15) +0.0749570*Z(26,25) +0.1578453*Z(40,41) +0.1697010*Z(60,59) 
-0.0074973*Z(82,83) -0.2285922*Z(110,109) -0.0261598*Z(140,141) +0.3019316*Z(176,175) -0.2923630*Z(214,215) 
+0.1586920*Z(258,257) -0.0595914*Z(304,305) +0.0169141*Z(356,355) -0.0038087*Z(410,411) +0.0007046*Z(470,469) 
-0.0001088*Z(532,533) +0.0000147*Z(600,599) -0.0000015*Z(670,671)

0.0000178 :  2:  -0.0030615*Z(6,5) -0.0254743*Z(12,13) -0.0850827*Z(24,23) -0.1618826*Z(38,39) 
-0.1591569*Z(58,57) +0.0257668*Z(80,81) +0.2284503*Z(108,107) +0.0073571*Z(138,139) -0.2934941*Z(174,173) 
+0.2969140*Z(212,213) -0.1652728*Z(256,255) +0.0632820*Z(302,303) -0.0182638*Z(354,353) +0.0041754*Z(408,409) 
-0.0007833*Z(468,467) +0.0001227*Z(530,531) -0.0000167*Z(598,597) +0.0000018*Z(668,669)

0.0000178 :  0:  -0.0030615*Z(1) -0.0254743*Z(4) -0.0850827*Z(11) -0.1618826*Z(22) -0.1591569*Z(37) 
+0.0257668*Z(56) +0.2284503*Z(79) +0.0073571*Z(106) -0.2934941*Z(137) +0.2969140*Z(172) -0.1652728*Z(211) 
+0.0632820*Z(254) -0.0182638*Z(301) +0.0041754*Z(352) -0.0007833*Z(407) +0.0001227*Z(466) -0.0000167*Z(529) 
+0.0000018*Z(596)
\end{verbatim}
\normalsize

\section{Corrigenda} \label{app.err} 
\subsection{Wang and Markey} \label{sec:WangMkey} 
Section IV in \cite{WangJOSA68} suffers from an inconsistent normalization
of integrals over the pupil. This becomes first apparent in (29), where
a division by $\pi D^2$ is missing on the right hand side.
(Equation numbers here in Appendix \ref{app.err} refer to the original papers \cite{WangJOSA68}
and \cite{FriedJOSA68}.)
The physical
units of $W$ in (2) and of $G$ in (28) are unity; the unit of $\phi$ is radians, which
means with (26), the $\beta$ also carry the unit radians; there is
a divisor missing with units of square meters to compensate the units of the
differential $d^2{\bf r}'$ in (29).
There are two
philosophies to correct this. The smaller amount of editing is needed if
(28) is re-normalized such that the factor $\pi D^2$ disappears---which 
resumes Fried's notation. Alternatively, the list of changes
consistent with keeping (28) is:
\begin{itemize}
\item
Divide the right hand side of (29) and the left hand side of (30) through
$\pi D^2$.
\item
Divide the middle term and the right hand side of (33) through
$\pi D^2$.
\item
Replace the factor $r$ in front of the integral in (34) by $r'$.
\item
Divide the left hand side of (35) through
$\pi D^2$.
\item
Multiply the right hand side of (37) by $\pi$.
This demonstrates that the eigenvalues in the $\lambda_{p,q}^2$
column of TABLE III are compatible with those in \cite[TABLE I]{FriedJOSA68}.
\end{itemize}
A second problem seems to have been induced by the mixed use of real-valued azimuthal functions
in (9) and complex-valued azimuthal functions in (32). To keep (32),
(43) is to be changed to
\begin{equation}
\int_0^{1/2}K_p^q(x)K_{p'}^q(x)x\, dx=\left\{
\begin{array}{l@{,\quad\:\mathrm{if}\:}l}
1/2 & p=p', \\
0 & p\neq p' .
\end{array}
\right.
\end{equation}

\subsection{Fried} 
Typographic corrections to \cite{FriedJOSA68} are:
\begin{itemize}
\item
$f_n$ in (9) ought read $f_n^*$.
\item
$\beta_{n'}$ in (16) ought read $\beta_{n'}^*$.
\item
$\langle \beta_n^*\beta_{n'}^*\rangle$ in (17) ought read
$\langle \beta_n^*\beta_{n'}\rangle$.
\item
${\cal G}_2(x)$ in (27c) ought simply read ${\cal G}_2$.
${\cal G}_4(x)$ in (27e) ought simply read ${\cal G}_4$.
\item
A minus sign is missing in front of ${\cal G}_0$ in (28).
\item
The factor ${\cal B}(x')$ should read
${\cal W}(x')$ in the integral in the
first line of (34).
\item
In the first line of (36a), the term $-2xx$ in the square root
ought read $-2xx'$.
\item
In (36b), a factor $x$ is missing in front of ${\cal G}_3(x')$, and
$\exp(i\theta)$ ought read $\exp(i\theta')$.
\item
In (36c), a minus sign is missing in front of the integral, and
$\exp(iq\theta)$ ought read $\exp(iq\theta')d\theta'$.
\item
The variables ${\cal B}_n$ ought be squared in (37a) and (37b).
\item
$(\frac{1}{4}\pi D^2)_{-1}$ in (40) ought read 
$(\frac{1}{4}\pi D^2)^{-1}$.
\end{itemize}

\subsection{Noll} 
Typographic clarifications to \cite{NollJOSA66} are
\begin{itemize}
\item
The name Wiener is misspelled after (18).
\item
$[\Gamma(P+1)/2]^2$ in the denominator of (20) is to be interpreted as
$[\Gamma\{(P+1)/2\}]^2$.
\item
the term $-2n$ in the exponent of the
second line of (25) ought read $-2m$.
\item
$\Gamma[(P+1/2]$ in the denominator of the unnumbered equation on page 211
is to be interpreted as
$\Gamma[(P+1)/2]$.
\end{itemize}

\bibliographystyle{apsrmp}
\bibliography{all}

\end{document}